# Can large-scale R&I funding stimulate post-crisis recovery growth? Evidence for Finland during COVID-19

**Authors:** Timo Mitze[1] and Teemu Makkonen[2*]

[1] University of Southern Denmark, Campusvej 55, 5230 Odense, Denmark, tmitze@sam.sdu.dk, https://orcid.org/0000-0003-3799-5200

[2] University of Eastern Finland, Yliopistokatu 2, 80101 Joensuu, Finland, teemu.makkonen@uef.fi, https://orcid.org/0000-0002-1065-1806

[*] Corresponding author

**Abstract:** The COVID-19 pandemic and subsequent public health restrictions led to a significant slump in economic activities around the globe. This slump has met by various policy actions to cushion the detrimental socio-economic consequences of the COVID-19 crisis and eventually bring the economy back on track. We provide an ex-ante evaluation of the effectiveness of a massive increase in R&I funding in Finland to stimulate post-crisis recovery growth through an increase in R&I activities of Finnish firms. We make use of the fact that novel R&I grants for firms in disruptive circumstances granted in 2020 were allocated through established R&I policy channels. This allows us to estimate the structural link between R&I funding and economic growth for Finnish NUTS-3 regions using pre-COVID-19 data. Estimates are then used to forecast regional recovery growth out of sample and to quantify the growth contribution of R&I funding. Depending on the chosen scenario, our forecasts point to a mean recovery growth rate of GDP between ~2–4% in 2021 after a decline of up to -2.5% in 2020. R&I funding constitutes a significant pillar of the recovery process with mean contributions in terms of GDP growth of between 0.4% and 1%.

## 1 Introduction

In 2020, the Finnish economy plummeted, along with the rest of the world, due to COVID-19. The pandemic was met with public health measures to restrict mobility and social contacts (Banholzer et al., 2021), which led to a decrease in overall consumer demand for many goods and services along with supply restrictions. Consequently, the volume of Finland's GDP fell (according to preliminary statistical data by Statistics Finland) by -1.8% in 2020. In 2021, the Finnish economy has started to recover as vaccinations have allowed to open up the society. According to the most recent figures published in September 2021, the Finnish employment rate has already recovered to pre-crisis level and the volume of GDP is growing again.[1] Depending on the forecast, the Finnish economy is expected to grow between 2.9–3.7% in 2021.[2]

In addition to the opening of the society, another important factor that contributes to this recovery process is massive public support given to the private sector: the amount of government subsidies to firms has grown significantly since 2020 compared to the pre-crisis situation. Besides standard instruments providing wage and cost subsidies, the Finnish government has also expanded its available support mechanisms for firms, among others, through the "Funding for business development in disruptive circumstances" program administered by Business Finland (BF), which allocates research and innovation (R&I) grants to firms with innovative ideas on tackling the detrimental effects of COVID-19 to their businesses. With an overall volume of 1,740M Euro in 2020 (compared to 570M in 2019) traditional and new BF funding channels are key fiscal policy measures used in Finland to stimulate economic activity during the COVID-19 crisis.

Against this background, it is the main research focus of this paper to better understand if and how the significant increase in R&I funding is likely to affect regional development trends in Finland. As no data for an *ex-post* assessment of policy effectiveness are yet available, we conduct an *ex-ante* evaluation to provide policy advice on this urgent matter. Specifically, we first estimate a structural regional economic growth model that captures the main transmission channels of BF funding prior to the COVID-19 outbreak and then use the estimated structural parameters for the pre-COVID crisis to forecast the growth stimulus of funding during the recovery phase after the initial COVID-19 shock. To identify the structural relationship between BF funding and economic growth, we exploit space-time differences in BF funding intensities at the level of 18 Finnish NUTS-3 regions over the period 1995–2018.

We then forecast regional GDP growth levels for the period 2019 to 2021 and compute the contribution of BF funding to the post-COVID recovery growth in 2021. For our baseline scenario, regional growth trajectories are found to be very close to the observed aggregate Finnish GDP slump in 2020 and match with the national (aggregate) forecasts of between 2.9–3.7% for 2021. That is, depending on the chosen forecast settings, our NUTS-3 level forecasts predict a maximum decline in regional GDP growth rates of up to -2.5% in 2020 (mean growth rate: -1%) and a mean recovery growth rate of between ~2-4% in 2021. Our forecasting results also show that the slump in 2020 and the post-crisis recovery growth in 2021 have particular spatial patterns. The results for recovery growth follow the commonly identified differences in regional development between the less developed eastern and northern and the more developed (core regions in the) southern and western parts of the country (Makkonen & Inkinen, 2015): the typical Finnish growth regions (e.g., the Finnish capital region of Uusimaa) are estimated to grow the fastest in 2021.

[1] For a recent update (in Finnish) on the state of the Finnish economy see (Accessed 13 October 2021): https://www.stat.fi/ajk/koronavirus/koronavirus-ajankohtaista-tilastotietoa/miten-vaikutukset-nakyvat-tilastoissa/talouden-tilannekuva

[2] The forecasts (in Finnish) can be found from (Accessed 13 October 2021):
Bank of Finland – https://www.eurojatalous.fi/fi/2021/3/ennuste-talous-ampaisee-vauhtiin-kun-pandemia-hellittaa/
ETLA Economic Research – https://www.etla.fi/ajankohtaista/etla-ennustaa-palveluvienti-ampaisee-ensi-vuonna-suurimmat-riskit-talouskasvulle-piilevat-pandemian-hallinnassa/
Ministry of Finance – https://julkaisut.valtioneuvosto.fi/handle/10024/163513
Pellervo Economic Research – https://www.ptt.fi/ennusteet/kansantalous-ja-asuntomarkkinat.html

Finally, our results predict that the significant increase in BF funding volumes constitutes a major pillar of this recovery process. Here our forecasts point to a mean growth contribution of changes in the BF funding intensity of between 0.4 and 1% in terms of regional GDP growth in 2021 (depending on the chosen recovery scenario). The positive contribution of BF funding is more important for those regions, outside the typical growth areas, with less endogenous regional growth factors. Effect size shrinks, though, if we assume that returns to “Funding for business development in disruptive circumstances” are lower than standard returns to publicly funded R&I activities.

The remainder of the paper is organized as follows: The next section summarizes the related literature on the link between R&I funding and regional growth. A specific focus is set on the role of R&I funding in times of crisis. Section 3 then presents institutional details about R&I funding in Finland. The section also explains how BF funding is used as major policy instrument to support the Finnish economy during the COVID-19 pandemic. While Section 4 presents the data, Section 5 outlines the estimation approach. This is followed by a presentation of the empirical results covering both in-sample predictions for the sample period 1995 to 2018 and out-of-sample forecasts until 2021 in Section 6. Section 7 concludes with a discussion of the policy relevance of our findings beyond the Finnish case.

## 2 R&I funding and regional growth

Innovation is considered as the key to economic development. There are naturally differences between the socio-economic conditions and institutional capacity of regions to turn innovation inputs such as public R&I funding into economic growth (Oughton et al., 2002). Still, the impact of R&I on regional economic growth has been found to be significant, particularly, for more developed regions in Northern Europe (Sterlacchini, 2008) but there is also some recent evidence for innovation-led regional development in emerging economies (Rodríguez-Pose & Villarreal Peralta, 2015). Therefore, public R&I funding is commonly regarded as a promising effort to achieve long-term goals like facilitating regional and economic growth, competitiveness and employment as well as tackling environmental and social problems (Acciai, 2021).

The need for public R&I rises from the averseness of private firms to invest on expensive and risky, but socially desirable, research due to knowledge spillovers that prevent them from fully cashing in on the potentially resultant new products and processes (Plank & Doplinger, 2018). On the contrary, public R&I funding programs are (or at least should be) designed to stimulate knowledge spillovers. As stated by Feldman and Kelley (2006: p. 1509): “R&I subsidies are an effective public policy instrument when knowledge spillovers exist". As explained by Veugelers (2021) public R&I funding can, however, also substitute (if the R&I project would have been carried out in any case without public funding) or crowd-out private R&I (by increasing the demand for and, thus also, the costs of R&I activities). Further, not all public R&I funding will lead to successful innovations nor subsequent economic outcomes.

Nonetheless, at least in Finland the positive link between public R&I funding and innovation is relatively clear: Torregrosa-Hetland et al. (2019) have demonstrated – based on innovation output data – that out of the identified 2 600 significant Finnish innovations, 35–55% were stimulated by the public sector. Further, a recent impact study (Fornaro et al., 2020) found clear evidence that public R&I funding increased the recipient firms' R&I intensity, R&D job creation, collaboration with external partners and productivity, while Piekkola (2007) found clear R&I spillover effects in terms of regional productivity and employment growth in Finland. In short, despite the obvious country and policy specific variations, there is abundant evidence suggesting that public R&I funding and regional development are linked (e.g., Becker, 2015; Boeing et al., 2022) through the mediating role that the funding can have on economic growth. For example:

1. Public sector R&I funding has been shown to lead to employment growth (Link & Scott, 2013) and particularly increase the number of R&D employees (Czarnitzki & Lopes-Bento, 2013).
2. Public sector R&I funding has been shown to increase, rather than substitute or crowd-out, the private sector's R&I activities and investments (Almus & Czarnitzki, 2003; Cin et al., 2017).
3. Public sector R&I funding is related to realized patenting by private firms (Haapanen et al., 2017; Czarnitzki & Hussinger, 2018).

Therefore, a major threat to innovation during a times of crisis is the slowdown of R&I funding and investments. Governments struggle with diminishing tax revenues and budgets and are compelled to implement short-term solutions rather than long-term development (Pellens et al., 2018), while firms might be forced to enter a survival mode postponing future-oriented R&I investments (Deschryvere et al., 2020). However, since innovation drives economic growth in the long run, governments are advised to find ways to increase, rather than decrease, their R&I funding in times of crisis (Makkonen, 2013).

As a specific reference to public R&I funding in times of crisis, Aristei et al. (2017) have indicated that public R&I funding has had at least a small positive impact on thwarting the reduction of firms' own R&I efforts in the aftermath of the economic crisis of 2008. Contrarily, Hud and Hussinger (2015), while reporting overall positive effect of R&I subsidies on SMEs' investment behavior, nonetheless found evidence supporting the remarks that public R&I funding crowded out private R&I during the economic crisis of 2008. As such, the evidence on the impacts of public R&I funding during times of crisis is still mixed. However, the evidence is leaning more towards positive outcomes (e.g., Brautzsch et al., 2015; Cruz-Castro et al., 2018). For example, based on data on OECD countries, Rehman et al. (2020: p. 349) conclude that in the case of the economic crisis of 2008: "public support to R&I is a good strategy for an economy to confront economic crisis effectively by increasing the technological innovation in the private sector".

While, the economic crisis of 2008 is very different from the contemporary crisis caused by the COVID-19 pandemic, the lessons learned from previous crises do, however, provide useful benchmarks when designating policy actions to mitigate the negative impacts of the contemporary one. The responses to tackling the COVID-19 pandemic indicate that governments have, at least partly, recognized that the best instruments for surviving and recovering from crises are related to supporting R&I (Bajaras et al., 2021). In fact, several countries have included R&I policies into their efforts in tackling the negative effects of the COVID-19 pandemic (Braunerhjelm, 2021). We will utilize one such country, Finland, as the empirical case in this paper.

## 3 The case of Business Finland

Business Finland (BF) is the Finnish government organization for innovation funding and trade, tourism and investment promotion. It operates under the Finnish Ministry of Employment and the Economy. It was formed in 2018 by merging its predecessors Tekes (the Finnish Funding Agency for Technology and Innovation) and Finpro (Finland Trade Promotion Organization) together. Business Finland is the main funding agency for research and technology development in Finland. As such it is a key intermediary between the government and innovative organizations (Inkinen & Suorsa, 2010) and a vital part of the Finnish innovation system (Ramstad, 2009).

Business Finland funds universities, research institutes, firms registered in Finland as well as public bodies. In the case of firms, R&I funding through Business Finland is normally given to encourage companies to improve their ability to develop and apply new technologies and to transform research-stage ideas into viable businesses via (a combination of) direct unconditional grants as well as soft and guaranteed low-interest or capital loans conditional on the success of the resulting business (Piekkola, 2007).

As explained by Takalo et al. (2013: p. 260), the public decision criteria of Business Finland are based on the project's estimated effects on the competitiveness of the applicant, the technology to be developed, the resources reserved for the project, the collaboration with other organizations and societal benefits. The applicants need to include the purpose and the budget of their intended R&I projects for which BF funding is needed together with the applied amount of funding (very large subsidies are rarely granted). The subsidy is granted as a share (depending on the size of the organization and the type of the project normally up to 40–65%) of the total R&I costs and mostly given after the R&I investments are made.[3]

In 2020, as a response to the COVID-19 pandemic, Business Finland launched a new funding program: "Funding for business development in disruptive circumstances".[4] The funding was channeled through two types of grants:

1. Preliminary funding for companies during business disruptions (max. €10,000 and 80% of the project's total costs, out of which up to 70% can be paid in advance): to be used for

[3] Information on the funding rules of Business Finland can be found from (Accessed 19 October): https://www.businessfinland.fi/en/for-finnish-customers/services/funding/research-and-development

[4] Information on funding rules for business support in disruptive circumstance can be found at (Accessed 13 October 2021): https://www.businessfinland.fi/en/for-finnish-customers/services/funding/disruptive-situations-funding

investigating and planning new business, alternative subcontracting chains, and ways to organize production during and after the disruption caused by the coronavirus

2. Development funding for companies during business disruptions (max. €100,000 and 80% of the project's total costs, out of which up to 70% can be paid in advance): to be used for carrying out development plans to improve the potential for success during and after the disruption caused by the coronavirus via the creation of new product- or production-related solutions

The funding criteria were uniform for firms across all Finnish regions and based on competition, but the funding program was specifically designed to fit the needs of SMEs. The maximum funding level can be considered quite modest compared to the regular funding given by Business Finland: in 2020, the average of granted funding per firm were 30,000€ smaller than in 2010–2019 (ca. 100,000€). The application period for the program ran in 2020. The program received almost 30,000 applications out of which circa 20,000 were funded. The total amount of granted funding via this instrument surmounted to 990M in 2020. The instrument was the largest of all COVID-19 support measures of the Finnish government that have, thus far, totaled up to 2,400M. The second largest support instrument, business cost support from the State Treasury amounted to 702M, while the rest (such as the remuneration for business restrictions in the restaurant business) of the measures have been significantly smaller.[5] Overall BF funding (established measures and the "Funding for business development in disruptive circumstances" program) reached a level of 1,740M in 2020 (compared to 570M in 2019).

## 4 Data and Stylized Facts

We utilize data collected at the NUTS-3 level in Finland for our empirical estimations. There are 19 NUTS-3 regions in Finland: 18 in mainland Finland and one covering the autonomous region of Åland. Due to missing data in some of the key variables utilized in this paper, Åland has been excluded from the analysis. The remaining NUTS-3 regions correspond to the second tier of the administrative regional division in Finland – state, regions (*maakunta*) and municipalities.

Descriptive statistics and exact definitions of the utilized variables at the NUTS-3 level are given in Table 1. While most of the data were gathered from the database (and archives) of Statistics Finland, patent data were derived from the OECD RegPat database and data on R&I funding were provided by Business Finland. The overall database covers the time period from 1995 (when Finland joined the EU) up to 2018 (the latest year available for the full set of utilized variables at the time of data collection in August 2021), with the exception of BF funding data which is available until 2020.

[5] For an overview (in Finnish) on government support instruments see (Accessed 13 October 2021): https://yle.fi/uutiset/3-12115110

Table 1: Variable descriptions and summary statistics

| Variable | Description | Obs. | Mean | S.D. | Min | Max |
|---|---|---|---|---|---|---|
| *GDP* | Regional GDP per employee (in Euro, at current prices) | 432 | 64477 | 13499 | 37334 | 100015 |
| *Δgdp* | Regional GDP growth rate (in %, annual based on GDP per employee) | 414 | 2.73 | 3.90 | -9.78 | 18.94 |
| *GFCF* | Regional gross fixed capital formation (in 1,000 Euro per employee) | 432 | 373.69 | 126.93 | 184.59 | 1310.44 |
| *EMPL* | Regional employment level (in 1,000) | 432 | 132.89 | 168.34 | 26.58 | 920.95 |
| $R\&D^{EXP\ BUS}$ | Regional business sector R&D expenditures (in % of GDP) | 432 | 0.96 | 1.00 | 0.09 | 5.34 |
| $R\&D^{EXP\ PUB}$ | Regional public sector R&D expenditures (in % of GDP) | 414 | 0.49 | 0.35 | 0.00 | 1.25 |
| $R\&D^{PER\ BUS}$ | Regional business sector R&D personnel (in % of total employment) | 432 | 1.06 | 0.76 | 0.18 | 3.50 |
| $R\&D^{PER\ PUB}$ | Regional public sector R&D personnel (in % of total employment) | 416 | 1.06 | 0.69 | 0.01 | 2.34 |
| *PATENT* | Regional patent applications (number per 1,000 employees) | 432 | 0.25 | 0.34 | 0.00 | 1.88 |
| *PATSTOCK* | Regional patent stock (number per employee) | 432 | 3.99 | 5.58 | 0.00 | 36.41 |
| *UNEMP* | Regional unemployment rate (in %) | 432 | 14.15 | 4.29 | 5.29 | 29.08 |
| *HIGHEDU* | Regional stock of highly educated employees (in % of total employment) | 432 | 9.06 | 3.04 | 4.34 | 21.51 |
| *BF FUNDS TOTAL* | Regional amount of Business Finland funding (in Mio. Euro; until 2018) | 432 | 24.47 | 51.33 | 0.16 | 316.74 |
| *BF* | Regional Business Finland funding (in Euro per employee; until 2018) | 432 | 122.58 | 85.43 | 6.01 | 409.94 |

*Notes:* Source information and further computational details are given in the main text.

Turning to some stylized facts of R&I funding through Business Finland, Panel A of Figure 1 shows the evolution of funding volumes over time. While overall funding volumes have moderately grown over the period 1995 to 2010, thereafter funding levels have mainly stayed constant or slightly declined around 2016 to 2018. More strikingly however, BF funding has more than tripled in 2020 compared to its average pre-COVID funding level. As outlined above, this increase can mainly be attributed to the new funding program under the heading "Funding for business development in disruptive circumstances". The latter amounts to approximately 57% of total R&I funding of Business Finland in 2020.

In addition, Panel B of Figure 1 (Boxplots) shows that the distribution of BF funding levels across regions (normalized by regional employment levels) also varies over time – with larger regional spreads being observed during times of crises, i.e., in the aftermath of the global economic crisis of 2008 and during the ongoing COVID-19 crisis. Panel A to C of Figure 2 present choropleth maps of funding receipt (again normalized by regional employment levels) for the sample years 2000 (Panel A) and 2020 (Panel B) together with the regional shares of the "Funding for business development in disruptive circumstances" program in Panel C (in percentage of total BF funding volumes in 2020).

Figure 1: Temporal evolution of BF funding volumes and regional spread of funding intensity

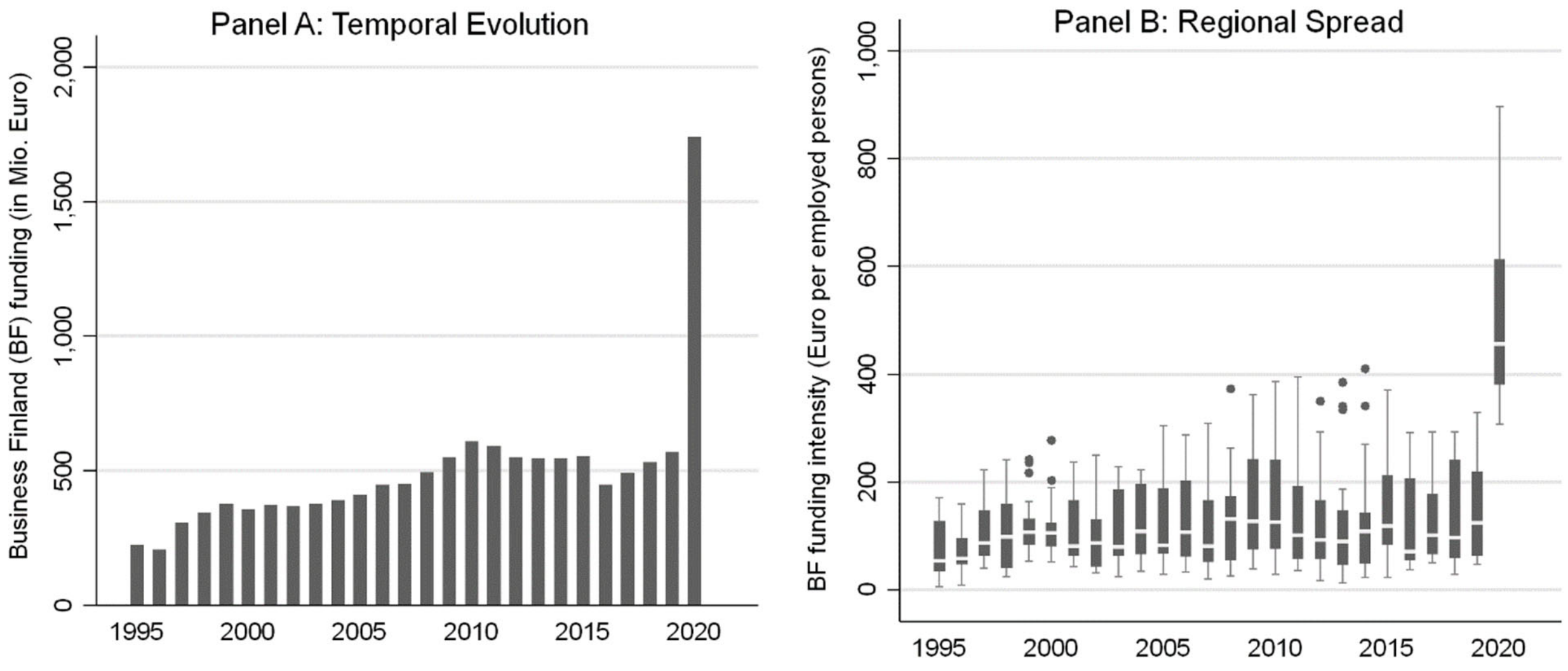

*Notes:* Own calculations based on data provided by Business Finland.

Figure 2: Spatial distribution of BF funding intensities and funding types

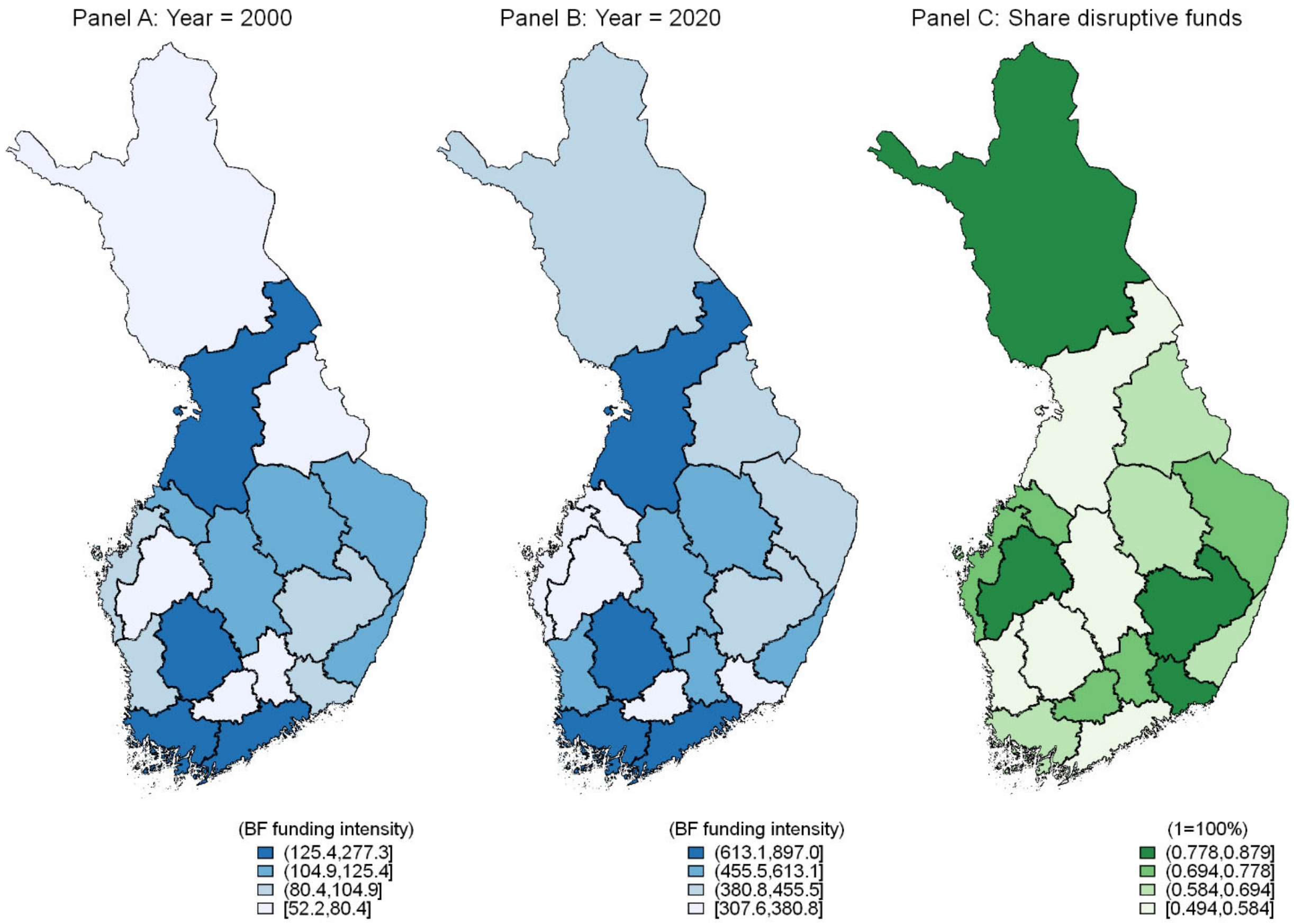

*Notes:* Own calculations based on data provided by Business Finland.

The regional distribution presented in Figure 2, which appears to be relatively stable over time, follows a general pattern, where the firms located in the most economically developed and populous regions of Finland have done consistently well in attracting BF funding. The top-ranking regions in terms of received BF funding include the regions of Uusimaa, Pirkanmaa, North Ostrobothnia and Southwest Finland. These four regions are centered around the largest Finnish cities of 1) Helsinki capital region (including Espoo and Vantaa), 2) Tampere, 3) Oulu and 4) Turku respectively (see Figure A1).

The result is in line with earlier evidence on regional development (Makkonen & Inkinen, 2015) and innovativeness (Valovirta et al., 2009) in Finland: the indicated regions (and cities) have done consistently (and historically) well in various benchmarking exercises based on socio-economic and innovation indicators. However, in terms of increases in BF funding intensities during the COVID-19 crisis, many of the smaller regions (in terms of population), such as the regions of Kainuu (+156.8%), South Savo (+153.9%) and Päijät-Häme (+153.1%), have significantly increased their funding intensities. When looking at the share of grants allocated through the new funding program (Panel C of Figure 2), we can see that funding plays a major role particularly in those regions that have not historically done so well in attracting BF funding.

## 5 Estimation Strategy

Our estimation/forecasting approach is organized as a two-stage process: First, we run a series of in-sample estimations for the pre-COVID sample period 1995 to 2018 (i.e., last available sample year for regional data at the time of this analysis). In this stage, we are mainly interested in consistently quantifying the structural relationship between regional R&I funding intensities by Business Finland and annual GDP growth rates, where the latter variable is chosen as broad regional performance measure. While BF funding is mainly directed towards enhancing the firms' R&I performance (as outlined above), we argue that an analysis of regional GDP growth rates (including R&I input factors) not only covers these underlying supply side mechanisms, but also accounts for potential demand-side effects of BF funding in the short run, e.g., in terms of using BF funding to hire research personnel, buy research and other equipment etc. To not over- or underestimate the overall funding effect, however, we also estimate auxiliary equations for (private and public) R&I inputs, i.e., regional R&D expenditure and patent rate, in which BF funding intensities enter as a determinant as well. These estimated links add to the total contribution of BF funding on regional GDP growth if R&I variables turn out to be a statistically significant determinant of regional growth.

To arrive at a consistent pre-COVID benchmark specification, in terms of estimator choice, we make use of recent advances in macro-econometric modelling for panel data together with the estimation of factor models utilizing (near-time) information from national time-series development for regional GDP forecasts (Kopoin et al., 2013; Lehmann & Wohlrabe, 2013). Specifically, available information on national GDP growth (or available growth predictions) are used as an essential scaling factor to predict regional GDP development in the course of the severe COVID-19 shocks throughout 2020 and 2021. In similar vein, we use other national macroeconomic data on employment and population levels as well as the aggregate unemployment rate to extrapolate the corresponding regional variable levels if they are needed for our growth model. Details are given below.

Formally, the main GDP growth equation to be estimated is specified as follows

$$\Delta gdp_{i,t} = \beta' \Delta \mathbf{x}_{i,t-m} + \lambda_i' \mathbf{F}_t + \phi \cdot ec_{i,t-n} + \mu_i + \varepsilon_{i,t}. \quad (1)$$

In eq.(1), $\Delta gdp_{i,t}$ measures the annual growth rate of GDP per employee (i.e., labor productivity) in region *i* at time *t* defined as $\Delta gdp_{i,t} = \log(GDP_{i,t})$- $\log(GDP_{i,t-1})$ and $\Delta \mathbf{x}_{i,t-m}$ is a vector of lagged regional short-run regional determinants (where *m* defines the lag length for these predetermined variables) including changes in regional gross fixed capital formation, changes in the regional unemployment rate, private and public R&D expenditures, patent applications and BF funding intensities. All variables in $\Delta \mathbf{x}_{i,t-m}$ used to estimate the growth model enter eq.(1) as logarithmic transformed values. The vector of coefficients $\beta$ accordingly measure the % response of regional GDP per employee growth to % changes in these regional short-run determinants as elasticities. As indicated by the lag structure in eq.(1), we rely on weak exogeneity (pre-determinedness) for effect identification. This clearly limits the interpretation of estimated effects with regard to their causal nature. We argue, however, that pre-determinedness together with a multi-factor panel estimation approach to avoid an omitted variable bias may be regarded as sufficient to obtain robust structural regression coefficients at the regional level.

Apart from the inclusion of regional short-run determinants, $\mathbf{F}_t$ captures common factors whose values vary across time but not across regions. In our default specification, we include national GDP growth as essential contemporaneous, common scaling factor for the development of regional GDP levels. This approach lends itself to the empirical literature on common correlated effects (CCE) estimators (Chudik et al., 2011). While the CCE literature typically employs cross-sectional averages of the dependent variable as a key proxy for unobserved common factors over time (Pesaran, 2006), the advantage of using the national GDP growth rate, particularly for out-of-sample forecasts, is that here observations are available beyond the maximum sample period at the regional level.

As we find that the correlation between both variables is very high ($\varrho = 0.86$), this indicates that both aggregate information and cross-sectional average can be used interchangeably without loss of estimation power. The coefficient (also referred to as factor loading in the CCE literature) that measure the response of regional GDP growth rates to changes in the national aggregate GDP growth rate can either be modelled to be constant across all regions ($\lambda$ for all *i*) or to vary across regions by taking the form $\lambda_i'$ (cf., Coakley et al., 2002; Pesaran, 2006). The latter heterogeneous coefficient approach acknowledges the fact the national-regional interrelations in GDP growth may be different for the set of included NUTS-3 region, for instance, mirroring different sectoral structures and, thus, local business cycle synchronization. We test if coefficient homogeneity holds in our data setting.

While the inclusion of short-run determinants ($\Delta\mathbf{x}_{i,t-m}$) in eq.(1) is expected to cover a significant share in the intra-regional variation of $\Delta gdp_{i,t}$, a pure short-run model of regional GDP growth may fail to account for relevant information contained in relationship between GDP levels and regional stocks of production factors such as the investment rate, the region's human capital endowment and further knowledge stocks. We make use of this long-run relationship in regional labor productivity levels and knowledge production factors in a single-equation cointegration framework (Engle & Granger, 1987; Phillips & Moon, 2000). In this context, the error correction term $ec_{i,t-n}$ captures deviations in this long-run relationship as

$$(2) \qquad ec_{i,t} = \log(GDP_{i,t}) - \left(\hat{\delta}'\mathbf{z}_{i,t} + \mu_i\right)$$

with $GDP_{i,t}$ being a measure of GDP per employee (labor productivity) in region *i* at time *t* and $\mathbf{z}_{i,t}$ is a vector of regional knowledge production factors. The coefficient vector $\hat{\delta}'$ captures the correlation between GDP and (knowledge) production factors in the long run, where the subscript "^" in eq.(2) indicates that these coefficients are estimated on the basis of a fixed effects panel model (FEM) for variables in log levels with $\mu_i$ being a vector of region-fixed effects.[6] In eq.(2), negative values of $ec_{i,t}$ indicate a mismatch between regional endowments and GDP per employee levels (a deviation from the estimated long-run co-integration path). We accordingly expect that an adjustment of GDP levels takes place over time, which implies that the regional economy will grow until the long-run cointegration relationship is restored. In this logic, the coefficient $\phi$ in eq.(1) will have a negative sign and that its magnitude measures the speed of adjustment of regional GDP levels towards the long-run cointegration relationship underlying eq.(2). The time index $t-n$ describes the chosen lag length for $ec_{i,t-n}$ in eq.(1) (see Table 2 for details how this is implemented).

[6] The set of regional determinants used for the long- ($\mathbf{z}$) and short-run estimation ($\Delta\mathbf{x}$) is allowed to differ from each other.

We include region-fixed effects in the short- and long-run equations to avoid an estimation bias arriving from unobserved time-constant factors at the regional level correlating with regional GDP level and growth rate differences. While eq.(2) is estimated by means of ordinary least squares (OLS), we additionally also check for potentially auto- and cross sectionally correlated errors in $\varepsilon_{i,t}$ by applying generalized least squares (GLS) estimation to eq.(1) next to the default FEM estimator.

As outlined above, we also estimate short-run auxiliary regressions for R&I inputs as

(3) $$\log(R\&D_{i,t}^{EXP\ BUS}) = \sum_{k=1}^{K} \psi_k \log(R\&D_{i,t-k}^{EXP\ BUS}) + \pi' \Delta \mathbf{x}_{i,t-i} + \mu_i + \epsilon_{i,t},$$

(4) $$\log(R\&D_{i,t}^{EXP\ PUB}) = \sum_{k=1}^{K} \tau_k \log(R\&D_{i,t-k}^{EXP\ PUB}) + \rho' \Delta \mathbf{x}_{i,t-i} + \mu_i + \omega_{i,t},$$

(5) $$\log(PATENT_{i,t}) = \sum_{k=1}^{K} \varphi_k \log(PATENT_{i,t-k}) + \theta' \Delta \mathbf{x}_{i,t-m} + \mu_i + u_{i,t},$$

to capture indirect effects of BF funding running through these R&I inputs. As no regional data are available yet for the out-of-sample forecasting period, short-run dynamics is captured as an autoregressive (AR) process with $\psi_k$, $\tau_k$ and $\varphi_k$ are the associated AR coefficients for lag *k*; $\theta$, $\pi$ and $\rho$ are coefficient vectors for the included short-run determinants $\Delta \mathbf{x}_{i,t-i}$ (see Table 2 for an overview of included variables in each equation). As in eq.(1) and eq.(2), $\mu_i$ are region-fixed effects and $u_{i,t}$, $\epsilon_{i,t}$ and $\omega_{i,t}$ denote the individual equations' error terms. As for the case of eq.(1), we estimate eq.(3) to eq.(5) by both OLS and GLS to account for auto- and cross sectional correlation patterns in the error term.

Based on the above-described set of estimates, we can compute the direct and indirect growth contribution of BF funding (in %) as

*Direct:* $\left(\hat{\beta}^{BF} \times \Delta\%BF\right)$

*Indirect:* $\left(\hat{\beta}^{PAT} \times \hat{\theta}^{BF} \times \Delta\%BF\right)$; $\left(\hat{\beta}^{R\&D\ EXP\ BUS} \times \hat{\pi}^{BF} \times \Delta\%BF\right)$; $\left(\hat{\beta}^{R\&D\ EXP\ PUB} \times \hat{\rho}^{BF} \times \Delta\%BF\right)$,

where $\Delta\%BF$ measures the percentage change in the BF funding intensity on a yearly base. To give an example: We calculate the direct GDP growth contribution of BF funding by taking the estimated coefficient $\hat{\beta}^{BF}$ (expressed as an elasticity, which measures the percentage change in GDP per employee growth for a 1% increase in R&I funding by Business Finland) and by multiplying it with the *de facto* percentage increase in the funding intensity for a given year, e.g., for 2020 to calculate the growth contribution in 2021. Similar calculations are done for the indirect contribution of BF funding running through the included R&I variables, where the BF contribution to changes in these variables is calculated on the basis of reduced-form coefficients.

In the second out-of-sample forecasting stage, we forecast regional GDP development for the years 2019 to 2021 and compute the contribution of BF funding to the post-COVID regional recovery. While the imposition of a given lag structure facilitates the computation of out-of-sample forecasts, we still need to extrapolate most of the included regressors (except BF funding for which we have data until 2020) in order to compute *h*-step ahead GDP growth forecasts on the basis of eq.(1). To minimize the need for data extrapolation as a source for forecast biases, the lag structure in the short-run growth equation is set to $m = 1$ and $n = 3$, which implies that we do not need to extrapolate time-series entering the long-run equation shown in eq.(2). Table 2 provides an overview of how we extrapolate the individual regressors included in eq.(1) in the out-of-sample forecast period 2019–2021.

Table 2: Extrapolation of regional variables entering the GDP growth equation

| **Variables in eq.(1)** | **Method of data extrapolation for out-of-sample forecasts** |
|---|---|
| $\Delta gfcf$ | Calculated as $\log(GFCF_t) - \log(GFCF_{t-1})$ based on predicted values of GFCF (see below) |
| $\log(GFCF)$ | AR(4) process (no national data available for out-of-sample period); prediction for 2019 and 2020 |
| $\Delta unemp$ | Calculated as $\log(UNEMP_t) - \log(UNEMP_{t-1})$ based on predicted values of UNEMP (see below) |
| $\log(UNEMP)$ | AR(4) process plus national development in unemployment rate; prediction for 2019 and 2020 |
| $\log(R\&D^{EXP\ BUS})$ | See eq.(3); *k*=3,4 and included short-run determinants: change in gross fixed capital formation, change in unemployment rate, BF funding intensity; prediction for 2019 and 2020 |
| $\log(R\&D^{EXP\ PUB})$ | See eq.(4) ; *k*=3,4 and included short-run determinants: change in gross fixed capital formation, change in unemployment rate, BF funding intensity; prediction for 2019 and 2020 |
| $\log(PATENT)$ | See eq.(5) ; *k*=3,4 and included short-run determinants: change in gross fixed capital formation, change in unemployment rate, BF funding intensity; prediction for 2019 and 2020 |
| $\log(BF)$ | BF Funds available until 2020; funding intensity for 2019 and 2020 calculated on the basis of predicted EMPL levels (see below) |
| $\log(EMPL)$ | AR(4) process plus national development in employment levels; prediction for 2019 and 2020 |
| $ec$ *(error correction term)* | Lag length set to *n* = 3 implies that no predictions of long-run variables beyond 2018 are needed to calculate GDP per employee growth rates until 2021 |
| $\Delta gdp$ (national) | Available data from Statistics Finland are taken for 2019 and 2020; for 2021 we expect that the rebound growth is of the same magnitude as the decline in 2020, i.e., 1.8% (baseline). Alternative scenarios also employ a lower rebound growth rate, e.g., a discount factor of *c*=0.5. See empirical results section for further details. |

*Notes:* AR(4) = Autoregressive panel model specification with a maximum of four time lags. All panel model specifications used to forecast variables shown in Table 2 include region-fixed effects.

## 6 Empirical Results

Table 3 presents the estimation results for regional GDP per employee growth in Finnish NUTS-3 regions as specified in eq.(1). While Column (1) reports the result of a benchmark specification excluding common factors and the error correction mechanism from a long-run GDP level equation, these additional factors are gradually added to the model in Columns (2) to (5). Most importantly, as the results table shows, in all specification we observe a statistically significant and positive conditional correlation between the BF funding intensity and regional GDP per employee growth. The range of estimated coefficient in Columns (2) to (5) of between 0.008 and 0.012 indicates that a doubling of BF funding intensities results in an increase in regional GDP per employee growth of 0.8-1.2%. While funding intensities typically do not vary in such large magnitude, the increase in BF funding intensities between 2019 and 2020 varied between 50% and 160%. This already indicates that the growth contribution of this policy instrument to fight the economic consequences of the COVID-19 may be considerable at the NUTS-3 level if constant growth returns to changes in the BF funding intensity are assumed. We turn to these out-of-sample predictions further below.

In terms of specification choice, two observations can be made from Table 3: First, criteria related to the goodness-of-fit of the in-sample GDP growth predictions (both $R^2$ and relative RMSE) significantly improve once we include a common factor structure (proxied by the national GDP development) and add the error correction mechanism estimated from a long-run GDP equation.[7] The latter has the expected negative sign indicating a medium-run adjustment process towards a long-run cointegration relationship specified in eq.(2). Second, we get statistical evidence that the association between regional and national GDP growth rates is heterogeneously distributed across the Finnish NUTS-3 regions (as indicated by the Wald test for coefficient equality of $\lambda_i$ reported in Table 2). Based on these observations, we argue that the most general and robust model specification is Column (5), which will be used as the basis for the out-of-sample forecasts.

One concern is, though, that the estimated link between regional BF funding intensities and GDP per employee growth rates may not be stable over time. In this case, our forecasts would overestimate the true funding effect on regional growth if the coefficient size declines over time. To test for coefficient stability, column (6) of Table 3 shows the estimation results for a sample-split regression, which only includes years from 2008 onwards. The subsample estimation results show no decline but a moderate increase in the coefficient for BF funding. Additionally, Figure 3 plots the estimation results of a variable coefficient approach for the full sample period, which captures potential time heterogeneity in the estimated BF funding coefficient by interacting the latter with separate time dummies covering three consecutive sample years (years before 2004 constitute the baseline period in this setup). As the results for the different estimators from column (1) to (5) in Table 3 show, coefficient size (except for model (1)) is very stable over time. Taken together, these robustness tests do not give any indication for in-sample parameter instability.

[7] The relative Root Mean Square Error (RMSE) is defined as the ratio of the model's RMSE and the corresponding RMSE of an autoregressive growth specification with a maximum of four lags (AR(4)).Relative RMSE values of below 1 thus indicate that the prediction error of the model is smaller than the one of the AR(4) benchmark specification.

Table 3: In-sample estimates of regional GDP growth model for Finnish NUTS-3 regions

| **Column** | **(1) OLS** | **(2) OLS** | **(3) OLS** | **(4) OLS** | **(5) GLS** | **(6) GLS** |
|---|---|---|---|---|---|---|
| **Dep. Var.:** | $\boldsymbol{\Delta gdp}$ | $\boldsymbol{\Delta gdp}$ | $\boldsymbol{\Delta gdp}$ | $\boldsymbol{\Delta gdp}$ | $\boldsymbol{\Delta gdp}$ | $\boldsymbol{\Delta gdp}$ |
| **Sample:** | **1998-2018** | **1998-2018** | **1998-2018** | **1998-2018** | **1998-2018** | **2008-2018** |
| $\Delta gfcf$ | -0.023*** | -0.006 | -0.005 | -0.006 | -0.009 | -0.006 |
| | (0.0074) | (0.0057) | (0.0062) | (0.0070) | (0.0056) | (0.0057) |
| $\Delta unemp$ | -0.036* | 0.060*** | 0.061*** | 0.058*** | 0.054*** | 0.062*** |
| | (0.0187) | (0.0167) | (0.0175) | (0.0170) | (0.0122) | (0.0097) |
| $\log(R\&D^{EXP\ BUS})$ | -0.033*** | -0.008 | -0.010 | -0.006 | -0.004 | -0.025*** |
| | (0.0070) | (0.0067) | (0.0078) | (0.0078) | (0.0056) | (0.0077) |
| $\log(R\&D^{EXP\ PUB})$ | 0.003 | 0.008*** | 0.009*** | 0.009*** | 0.009*** | 0.009*** |
| | (0.0023) | (0.0019) | (0.0023) | (0.0025) | (0.0022) | (0.0033) |
| $\log(PATENT)$ | -0.003 | -0.003* | -0.003* | -0.003** | -0.003** | -0.003* |
| | (0.0024) | (0.0016) | (0.0015) | (0.0013) | (0.0013) | (0.0015) |
| $\log(BF)$ | 0.016*** | 0.012*** | 0.012*** | 0.012*** | 0.008** | 0.009** |
| | (0.0047) | (0.0031) | (0.0032) | (0.0040) | (0.0038) | (0.0039) |
| $ec$ (*error correction term*) | | | | -0.150*** | -0.167*** | -0.099*** |
| | | | | (0.0296) | (0.0294) | (0.0361) |
| Obs. (*N*=Regions, *T*=Years) | 378 (*N*=18, *T*=21) | 378 (*N*=18, *T*=21) | 378 (*N*=18, *T*=21) | 378 (*N*=18, *T*=21) | 378 (*N*=18, *T*=21) | 198 (*N*=18, *T*=11) |
| Region-fixed effects | YES | YES | YES | YES | YES | YES |
| National GDP growth (common factor) | NO | YES (common coefficient $\lambda$) | YES (heterogenous coefficients $\lambda_i$) | YES (heterogenous coefficients $\lambda_i$) | YES (heterogenous coefficients $\lambda_i$) | YES (heterogenous coefficients $\lambda_i$) |
| Wald $\chi^2$test for equal $\lambda_i$ | | | 45.01*** | 495.33*** | 100.58*** | 87.31*** |
| Goodness-of-fit ($R^2$) | 0.10 | 0.40 | 0.42 | 0.45 | 0.45 | 0.38 |
| Relative RMSE | 0.94 | 0.77 | 0.77 | 0.78 | 0.76 | — |

*Notes:* * p < 0.10, ** p < 0.05, *** p < 0.01. Robust standard errors are given in brackets. GLS estimates control for the presence of AR(1) autocorrelation within and heteroskedasticity across panels. $R^2$ is calculated as squared correlation between predict regional growth per employee rate and its observed counterpart; RMSE is defined in text.

Figure 3: Test for parameter stability in the link between BF funding and regional GDP growth

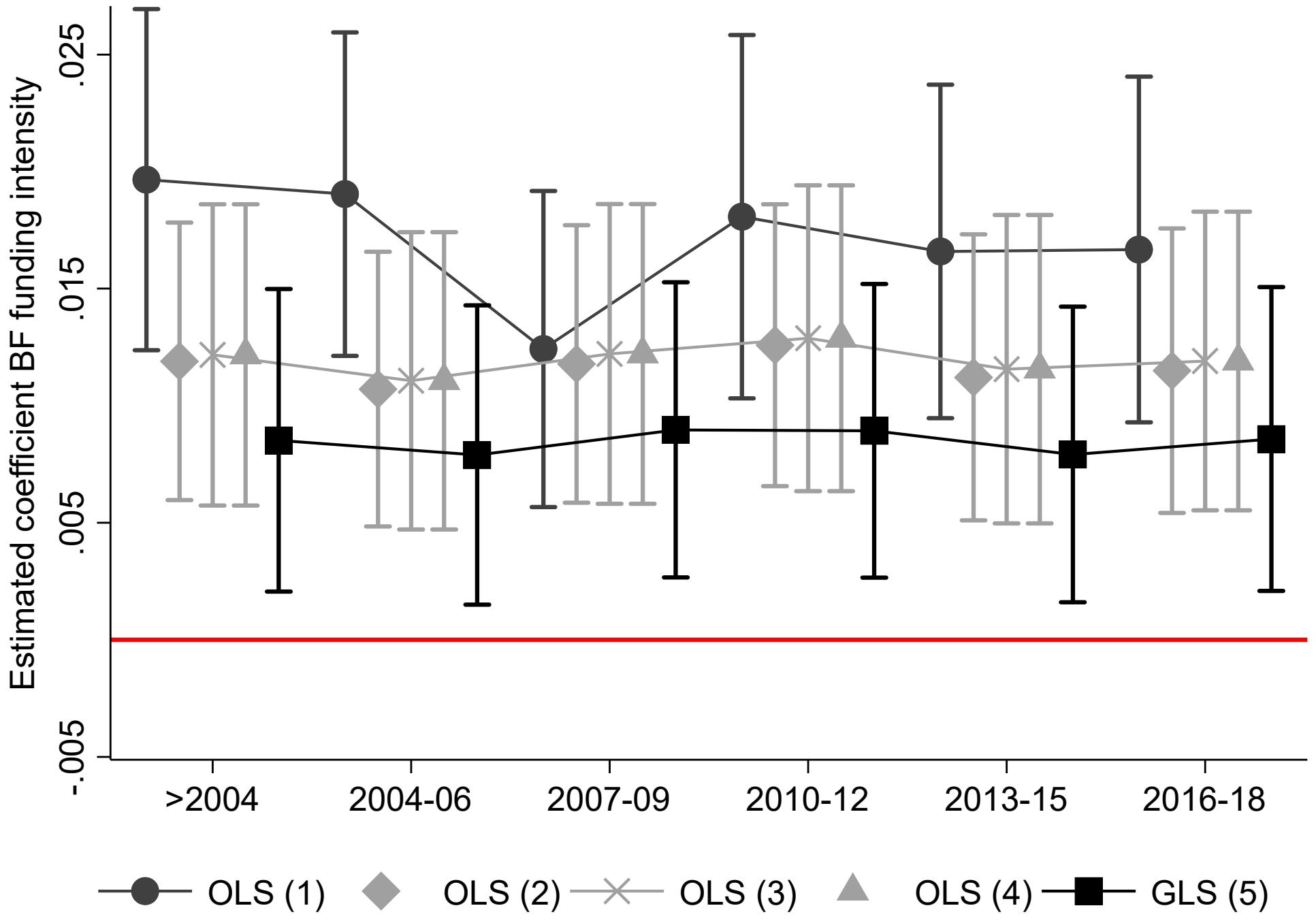

*Notes:* Estimates based on columns (1) to (5) in Table 3 with time-varying coefficients for the coefficient of the BF funding intensity pooled over a three-year period. Vertical lines indicate 90% confidence intervals.

Regarding the other included regional determinants, we find for most specifications in Table 3 that business sector R&D expenditures and the patent rate are either not or weakly negatively correlated with GDP growth. While this is likely due to the very volatile nature of patent applications at the Finnish NUTS-3 level and due to the fact, that after the financial crisis of 2008, private R&D expenditures have increased steadily only in the region of Uusimaa, we find that the volume of public sector R&D expenditure flows at the regional level is positively correlated with the regional GDP development in the short run. Moreover, in the long run, we also find a positive relationship between the region's patent stock, business sector R&D personnel, the physical capital investments, highly skilled labor, and regional growth. These positive long-run correlations are in line with most empirical contributions on growth and development at the Finnish national (Pohjola, 2017) and regional levels (Makkonen & Inkinen, 2015). The underlying structural parameters of the long-run GDP equation (with standard errors given in squared brackets) are estimated as

$$\log(GDP_{i,t}) = \underset{[0.0099]}{0.041} \cdot \log\left(GFCF_{i,t}\right) + \underset{[0.0159]}{0.716} \cdot \log\left(HIGHEDU_{i,t}\right) - \underset{[0.0130]}{0.182} \cdot \log\left(UNEMP_{i,t}\right)$$

$$+\underset{[0.0045]}{0.017} \cdot \log\left(PATSTOCK_{i,t}\right) + \underset{[0.0133]}{0.067} \cdot \log\left(R\&D^{EMP\ BUS}\right) + \underset{[0.0054]}{0.006} \cdot \log\left(R\&D^{EMP\ PUBL}\right)$$

$R^2$ = 0.63, Phillips-Perron *t*-statistic = -3.91$^{***}$ ($H_0$: No cointegration)

We also find evidence for a significant cointegration relationship between variables included in the long-run equation in the reported test statistic for a Phillips-Perron *t*-test with the null hypothesis of no cointegration among regions (Pedroni, 1999; 2004).

In addition to the direct relationship between BF funding and regional growth, Table 4 shows that this type of R&I funding is also associated with positive conditional correlations with the included R&I variables (except the very volatile patent rate) in eq.(1). This finding is in line with the earlier literature on public R&I funding as already discussed in Section 2: BF funding affects private R&I investments positively. While these indirect transmission channels thus suggest that BF funding further increases regional GDP growth through a technological upgrading mechanism, in the short run, we have to be aware that the link between some of these R&I inputs, i.e., private-sector R&D expenditures and the region's patent rate, was estimated to be absent and/or negative. This leaves the direction of the indirect growth contribution of BF funding unclear *ex ante*.

Table 4: In-sample estimates of auxiliary R&I equations

| **Column** | **(1)** | **(2)** | **(3)** |
|---|---|---|---|
| **Dep. Var.:** | $\log(R\&D^{EXP\ BUS})$ | $\log(R\&D^{EXP\ PUB})$ | $\log(PATENT)$ |
| **Sample:** | **1998-2018** | **1998-2018** | **1998-2018** |
| $AR$ coefficients (jointly) | $0.406^{***}$ | $0.445^{***}$ | $0.103^{*}$ |
| | (0.0394) | (0.0297) | (0.0598) |
| $\Delta gfcf$ | 0.004 | -0.019 | 0.113 |
| | (0.0209) | (0.0190) | (0.0777) |
| $\Delta unemp$ | -0.064 | $0.172^{***}$ | $0.373^{**}$ |
| | (0.0511) | (0.0438) | (0.1475) |
| $\log(BF)$ | $0.048^{**}$ | $0.058^{***}$ | -0.001 |
| | (0.0210) | (0.0215) | (0.0700) |
| Obs. (*N*=Regions, *T*=Years) | 360 (*N*=18, *T*=20) | 360 (*N*=18, *T*=20) | 360 (*N*=18, *T*=20) |
| Region-fixed effects | YES | YES | YES |

*Notes:* * p < 0.10, ** p < 0.05, *** p < 0.01. Standard errors are given in brackets. All estimates on the basis of GLS, which controls for the presence of AR(1) autocorrelation within and heteroskedasticity across panels.

Based on the in-sample estimation results shown in Tables 3–4, together with the forecast setup outlined in Table 2, we are finally able to compute and plot out-of-sample forecasts for regional GDP growth in Finland beyond the latest available regional observation in 2018. Reduced-form results for the in- and out-of-sample forecast period are shown in Figure 4. The figure plots the observed mean GDP per employee growth rate together with its mean prediction growth across the 18 Finnish NUTS-3 regions and a 95% forecast range based on +/- 2 standard deviations of the individual regional forecasts (see Figure A2 for the individual forecasts for each NUTS-3 region). These baseline forecasts are produced for the following scenario:

1) a national rebound growth rate of 1.8% in 2021 (after -1.8% in 2020),

2) constant GDP growth returns to BF funding.

Alternative scenarios are presented further below.

Figure 4: In- and out-of-sample forecasts of reginal GDP growth rates in Finnish NUTS-3 regions

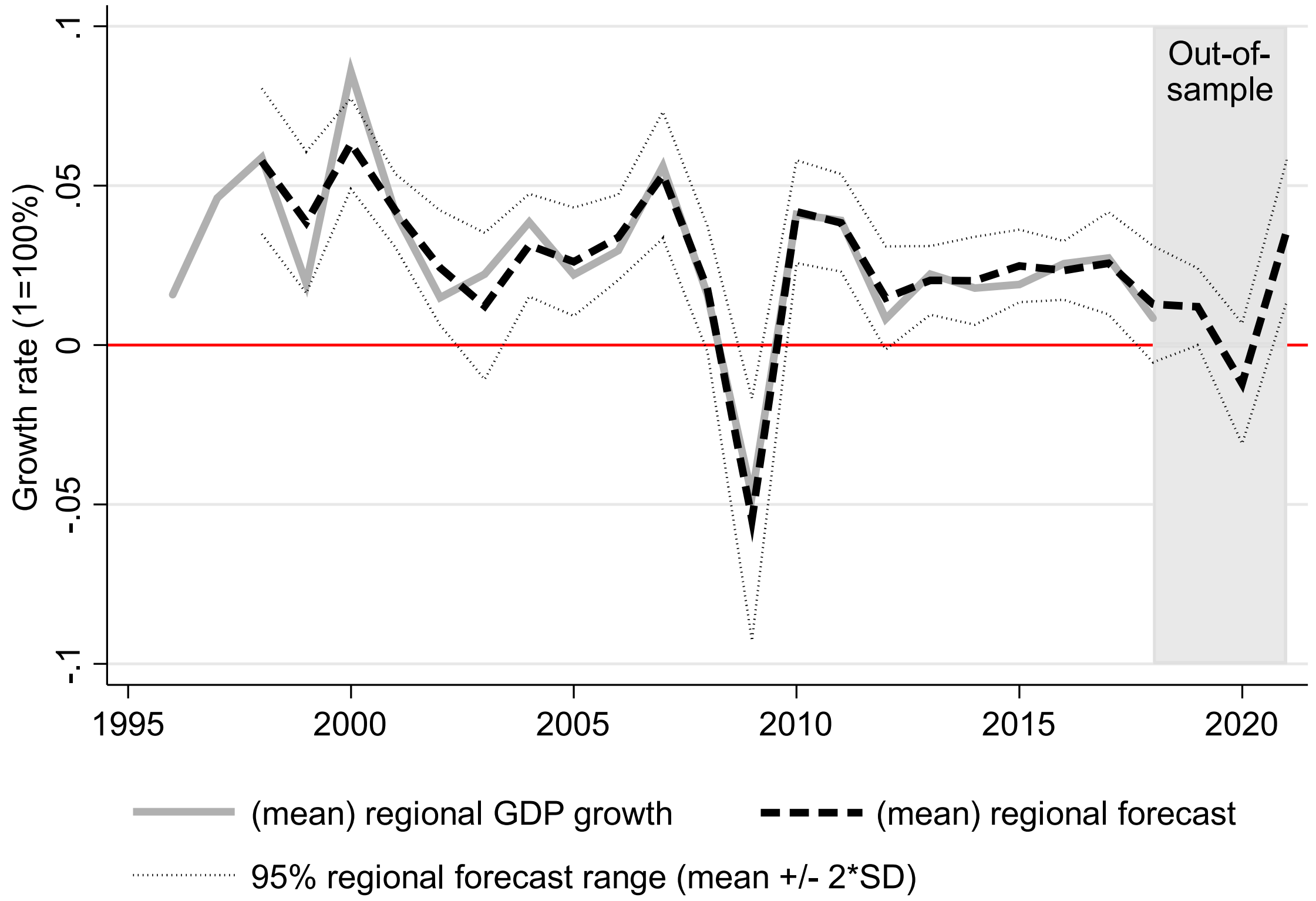

*Notes:* Own estimates based on parameters reported in Column (5) of Table 3 and forecast setup described in Table 2 and the text. National GDP growth in Finland in 2021 is assumed to be 1.8%.

If we compare the mean GDP growth rate (black solid line) with the in-sample prediction until 2018 (grey dashed line), Figure 4 shows that our mean forecast accurately matches the de facto development (as also indicated by the RSME<1 in Table 3). Only during the period 2000–2004 we observe a tendency of over-/underprediction of growth rates. However, in all cases the predicted model correctly identifies turning points in the data. Importantly, the model prediction does a particularly good job to match mean regional development during the economic crisis of 2008, which can be seen as an in-sample validation for the model's ability to capture the 2020 slump during the COVID-19 crisis. This is shown in the right area of the figure depicting the out-of-sample forecast interval. With regard to the magnitude of the crisis response of regional GDP levels we find a decline in regional GDP growth rates of up to -2.5% in 2020 (with a mean decline of ~-1%) and recovery growth rates of between 2 and 6% in 2021 (with a mean growth rate of ~4%). This prediction is slightly more optimistic but comes close to aggregate forecasts for the Finnish economy ranging from 2.9–3.7% (see Section 1). One should note, though, that the mean growth rate reported here puts an equal weight on each region and, hence, does not take the region's share in national GDP into account. Panels A to C of Figure 5 show how the different components of the short-run growth equation contribute to the forecasted post-crisis recovery growth across Finnish NUTS-3 regions.

Figure 5: Spatial distribution of main components of GDP growth in 2021 (baseline scenario)

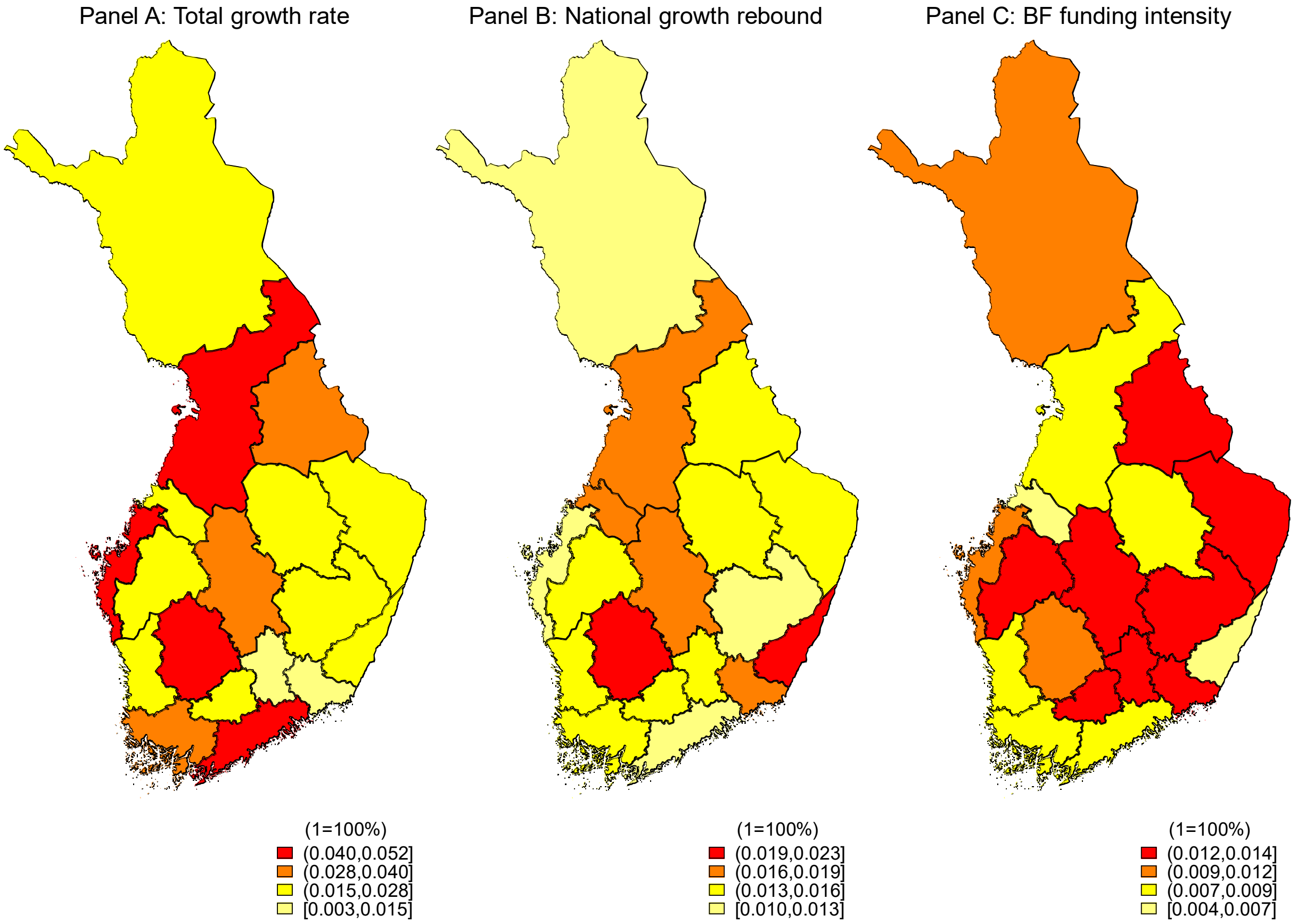

*Notes:* Own estimates based on parameters reported in Column (5) of Table 3 and forecast setup described in Table 2. The estimated vector of fixed effects has been ignored in the decomposition of GDP growth rates. National GDP growth in Finland in 2021 is assumed to be 1.8%.

As Panel A of Figure 5 shows, the largest total growth rate in 2021 is predicated for regions that are typically considered as the regional "growth engines" of Finland: Uusimaa, Pirkanmaa and North Ostrobothnia. Additionally, regions whose economies have been reported to experience significant growth, e.g., Ostrobothnia, and those whose economies have been reported to decrease, e.g., Kymenlaakso, before the COVID-19 crisis (European Commission, 2020) are predicted to experience, respectively, higher and lower recovery than on average. As such, the COVID-19 pandemic seems to have had a relatively small impact on the overall picture of spatial development in Finland. Panel B of Figure 5 shows the contribution of the predicted national GDP growth rate on regional GDP per employee growth rates running through $\lambda_i'\mathbf{F}_t$. Here, national-regional linkages account for between 1% and 2.2% of the predicted regional growth performance in 2021. As the panels show – and in line with earlier evidence (Pekkala, 2000) – regions do not grow at the same pace as the aggregate economy. In our forecasts the regions of Pirkanmaa and South Karelia are the most aligned with the national growth trend.

Panel C of Figure 5 plots the growth contribution of the massive increase in BF funding intensities in 2020, which are estimated to have a significant impact on regional GDP per employee growth in 2021. The estimated contribution of BF funding varies between 0.4% and 1.4% in terms of regional GDP per employee growth across Finnish NUTS-3 regions (mean growth contribution: ~1%).[8] Thus, on average, we find that the significant increase of BF funding in 2020 constitutes a major pillar of this recovery process as roughly one fourth of the overall regional GDP growth rate can be attributed to the increase in the BF funding intensity at the regional level (annual BF funding intensities increased by between 50% and 160% in 2020).

Especially many of the smaller (in terms of population) Finnish regions show significant positive growth due to BF funding. Regions – such as Kainuu, South Savo and Päijät-Häme – that have increased their BF funding intensity the most are the ones with the highest estimated BF funding contribution to their GDP growth rates. Additionally, in some cases the overall growth rate would even be negative, if not accounting for BF funding (and the national rebound effect discussed above). Contrarily in regions (including, for example, the already well-off Uusimaa) where the BF funding intensity has not grown as significantly, the contribution of BF funding on their estimated GDP per employee growth rate is more modest. Thus, the positive effect of BF funding seems to be more important for regions with less endogenous regional growth factors.

As pointed out above, the size of individual growth contributions and the overall GDP growth forecast depend on two key parameter settings in the baseline scenario. One is that the link between BF funding and regional growth (the return to BF funding) is considered constant even for significant higher funding levels and that this constant link also holds for the "Funding for business development in disruptive circumstances" – although the program deviates in some ways from the traditional BF funding (see Section 3). Earlier work on the effectiveness of regional funding for GDP growth has pointed to a potential maximum funding intensity after which no additional growth impulse can be observed (Mitze et al., 2015) pointing to a maximum absorptive capacity of regions to translate R&I (and other types of) inputs into outputs (Oughton et al., 2002). To accommodate this aspect, we run a series of forecasts that assume constant returns for traditional R&I funding from Business Finland but decreasing returns for the "Funding for business development in disruptive circumstances" program. We model decreasing returns by multiplying the estimated elasticity from Column 5 of Table 3 with a discount factor *r* (i.e., we choose values for *r* as *r* = 0.8; 0.7; 0.6). In other words, the direct growth contribution of BF funding is calculated as

*Direct:* $\left(\hat{\beta}^{BF} \times \Delta\%BF_{traditional}\right) + \left(r \cdot \hat{\beta}^{BF} \times \Delta\%BF_{disruptive}\right),$

where $\Delta\%BF_{traditional}$ measures the change in traditional BF funding intensity and $\Delta\%BF_{disruptive}$ the change in the intensity of "Funding for business development in disruptive circumstances" on a yearly base. We assume that the indirect BF funding channel is unchanged. Resulting forecasts for the BF funding contribution to regional GDP per employee growth in 2021 are plotted in Figure 6.

[8] It should be noted that the overall growth contribution is almost entirely driven by the direct BF contribution running through eq.(1), while the indirect contribution through eq.(3) to eq(5) only accounts for 7% of the total contribution.

Figure 6: Growth contribution of BF funding under varying degrees of R&I returns

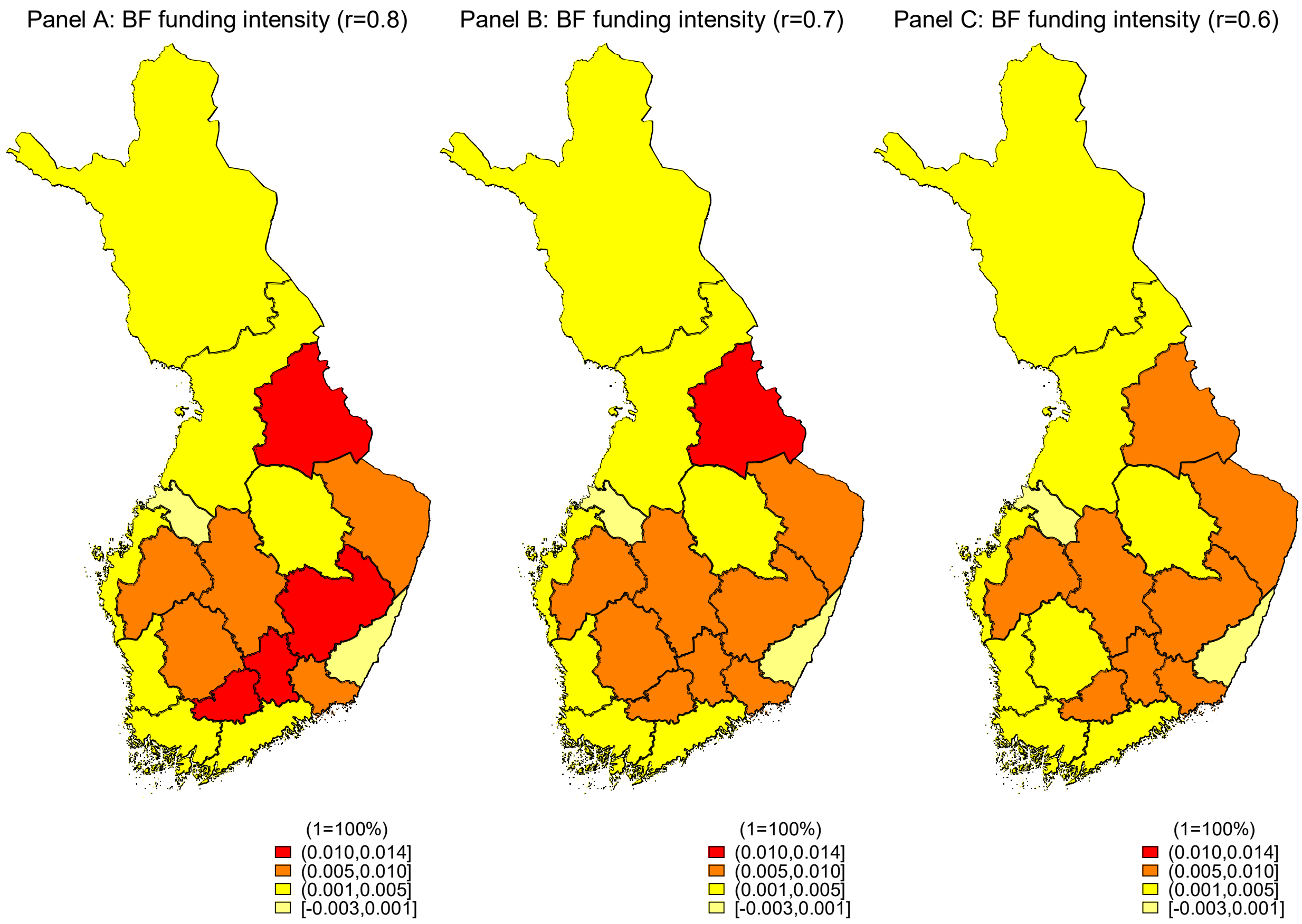

*Notes:* Own estimates based on parameters reported in Column (5) of Table 3 and varying discount factors *r* for the new program "Funding for business development in disruptive circumstances"; see text for further explanations.

As Figure 6 shows, larger discount factors generally translate into a lower average growth contribution of between 0.4 to 0.7% in terms of GDP growth (compared to an approx. 1% mean contribution in the baseline scenario) and also result in smaller inter-regional growth contribution differences. Particularly those regions that have significantly increased their overall BF funding intensity in 2020 through the new "Funding for business development in disruptive circumstances" program (see Panel C in Figure 2) are now observed to have smaller BF growth contributions *vis-à-vis* the rest of the country. Moreover, regions which – at the aggregate level – mainly substitute traditional BF funding in favor of grants from the new program without significantly increasing their overall BF funding intensities are now predicted to have a close to zero or even slightly negative BF growth contribution as shown in Figure 6. Taken together, we argue that a moderate discount factor of *r*=0.7 may be regarded as a reasonable choice given that the average BF funding volume per recipient firm is approx. 30% smaller in the new "Funding for business development in disruptive circumstances" program compared to the traditional BF funding channel. Smaller or lacking scale effects at the individual firm level may then accordingly translate into decreasing returns to BF funding at the regional level.

We use this setting to compute a more conservative (lower bound) growth scenario, which additionally assumes a smaller national GDP growth impulse for regional recovery trends (running through the included common factor structure). Specifically, we discount the 2021 prediction for national GDP growth (1.8%) by a factor of $c$ = 0.5 or, in other words, we halve the national growth impulse in this conservative scenario. The resulting spatial distribution of the main components of GDP growth in 2021 are shown in Figure 7. As the results in Panel A of Figure 7 show, this lower-bound scenario predicts a moderate mean regional GDP per employee growth rate of ~2% in 2021 compared to the ~4% mean recovery growth rate in the baseline scenario. While some regions are still predicted to have annual growth rates between 3% and 4%, at the lower end of the regional growth distribution, in two cases, this scenario would now predict that regions continue to shrink in 2021.

Figure 7: Spatial distribution of main components of GDP growth in 2021 (conservative scenario)

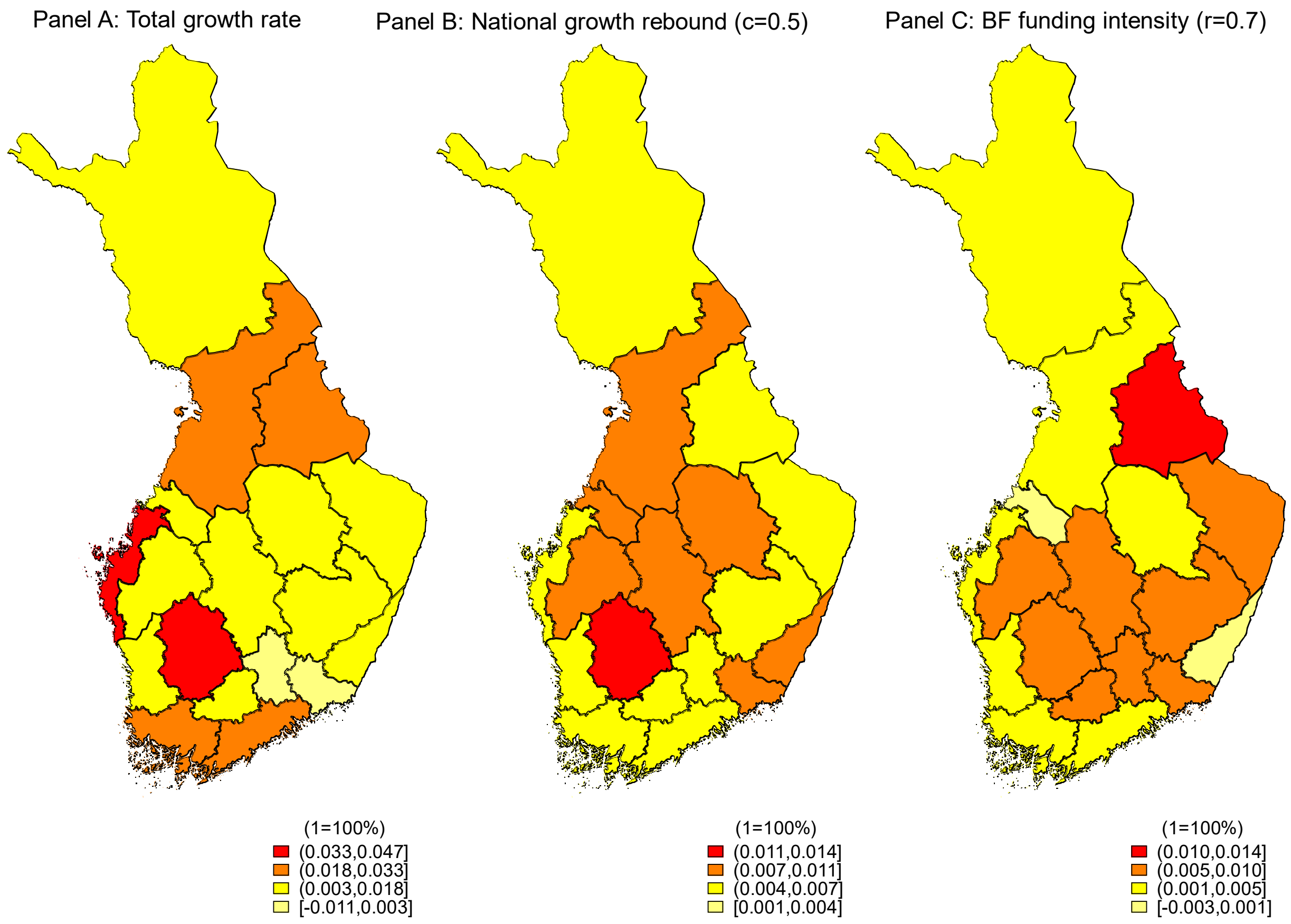

*Notes:* Own estimates based on parameters reported in Column (5) of Table 3 and forecast setup described in Table 2. The estimated vector of fixed effects has been ignored in the decomposition of GDP growth rates. The growth contribution of BF funding is calculated for a discount factor of $r$=0.7. For the national GDP growth impulse, we use a discount factor of c=0.5 compared to the baseline scenario; see text for further explanations.

## 7 Discussion and Conclusion

Fighting the detrimental socio-economic consequences of the ongoing COVID-19 crisis is a major challenge for policy makers. Many countries have started to experiment with different demand and supply-side oriented support schemes to assist individual workers and firms to survive in troubled times and to foster post-crisis recovery growth. One particular focus of the Finnish government was to provide public support to pursue R&I activities: BF funding increased from 570M in 2019 to 1,740M in 2020. In this paper, we have collected empirical evidence to make first, scientifically grounded statements about likely effectiveness of this policy plan.

Since regional data for a comprehensive ex-post evaluation of BF funding effectiveness are not readily available yet, we have adopted a two-stage ex-ante evaluation design. In a first step, we have estimated the empirical link between R&I funding through Business Finland and regional GDP growth for Finnish NUTS-3 regions. Our results for the pre-COVID sample period 1995–2018 point to a robust, positive correlation between the BF funding intensity and annual GDP growth. We additionally find that BF funding also affects the level of other R&I variables such as private and public R&D expenditure levels positively (although the link between these variables and GDP growth is found to be weaker in the short than in the long run). Based on these in-sample prediction, we have then forecasted regional GDP growth rates out-of-sample until 2021. An inspection of the forecast quality has shown that our regional forecasting system accurately predicts i) the COVID-related economic slump in 2020 and ii) the predicted national post-crisis recovery growth rate.

With regard to the role of R&I funding by Business Finland, we find that the massive increase in expenditure volumes is pivotal for this recovery growth rate. Based on our structural growth model estimates we find an overall growth contribution attributable of increased BF funding through Business Finland of between 0.4% and 1% of regional GDP per employee growth for the individual Finnish NUTS-3 regions. While we have carefully assessed the statistical significance of this result in the paper, one remaining question that still needs to be answered is: *How realistic is the estimated magnitude of the BF funding contribution to regional GDP growth?* To provide an answer to this question, Table 5 summarizes key information on our two main forecast scenarios together with information on the relative financial importance of BF financial inflows into a NUTS-3 region as share of its regional GDP.

Table 5: Relation of forecasted BF contribution and financial importance of funding for regional GDP

| | Mean | S.D. | Min. | Max. |
|---|---|---|---|---|
| Share of BF funding in regional GDP (in %, in-sample period, 1998-2018) | 0.18 | 0.11 | 0.02 | 0.57 |
| Share of BF funding (2020) in regional GDP (2018) (in %) | 0.60 | 0.16 | 0.36 | 0.94 |
| Predicted BF contribution to GDP growth rate of GDP (in %, baseline scenario) | 1.04 | 0.29 | 0.43 | 1.40 |
| Predicted BF contribution to GDP growth rate of (in %, conservative scenario with *r*=0.7) | 0.52 | 0.34 | -0.20 | 1.02 |

*Notes:* Predicted BF contributions to regional GDP growth rate are based on forecasts shown in Figure 5 and Figure 7.

The motivation for comparing the regional financial importance of BF funding with our forecasting results is that a predicted growth contribution of between 0.4% and 1% would be unrealistic if financial flows into a region are simply too small to expect sizable effects on regional GDP. However, as Table 5 shows, the financial importance of BF funding as share of regional GDP – in fact – lends further support to our forecasts. That is, while during the pre-COVID period the average financial importance of BF funding in regional GDP was indeed only about 0.2% (with a regional range of 0.02% to 0.6%), we observe a considerably higher average share of 0.6% (with a range of 0.36% to 0.94%) once we relate BF funding levels in 2020 (i.e., under COVID-19) to 2018 regional GDP levels (last in-sample observation). The latter regional range, in fact, comes close to our lower and upper bound effects for the growth contribution of BF funding in our different forecast scenarios. We thus argue that the predicted growth contribution of R&I funding through Business Finland is not only robust from a statistical perspective but also appears to be plausible from an economic perspective.

The results of this paper, thus, support the view that in addition to business cost support, governments should also invest on R&I as a means to recover from crises even in times of budgetary constraints. While short-term business cost support for firm survival has previously been found to be important when a crisis hits an economy (National Audit Office of Finland, 2021), the situation may change if the crisis lingers (Vihriälä et al., 2020): it can hinder market exit and thus lead to the misallocation of public funding towards supporting unviable firms. Or to put it differently: surviving a crisis also requires governments to design policies that encourage innovation (Acemoglu, 2009).

As far as it can be said from an ex-ante perspective, the Business Finland administered "Funding for business development in disruptive circumstances" seems to be a successful example of such policy implements. The emphasis given to the innovative aspects of the proposed projects rather than just on firm survival (Fornaro et al., 2020) in the selection process has arguably contributed to the success of this particular BF funding instrument. The granted funding per firm was designed to remain rather modest (either €10,000 or €100,000), which has allowed the funding to be spread amongst over 20,000 firms of which the majority were SMEs. In this regard, the fact that Business Finland covers 80% of the costs out of which 70% can be paid in advance can be argued to be a definite improvement for SMEs. This is because, even favorable R&I policies can be inefficient for small firms, if they need to cover the costs up-front, which might lead to cash-flow problems (particularly) in times of crisis (Harris et al., 2020).

Obviously, our finding of a positive growth contribution of R&I funding through Business Finland is conditional on the stability of the link between BF funding and regional growth during the COVID-19 crisis. If – instead – there is a maximum absorptive capacity of regions to translate publicly funded R&I inputs into outputs (Mitze et al., 2015), then the growth returns to BF funding may decrease with increasing funding intensities. The latter may also happen if the regional economy's absorptive capacity is lower due to the overall economic slump. We have accommodated this more conservative view in alternative forecasting scenarios, which point to a mean growth contribution of BF funding of between 0.4% and 0.7% in terms of GDP per employee growth (compared to ~1% in the baseline scenario). These scenarios assume that effect size declines in the new "Funding for business development in disruptive circumstances" program with smaller funding volumes per recipient firms compared to the traditional BF funding. Together with a more conservative projection about the size of the national growth impulse for regional recovery growth, we obtain a lower-bound forecast for the overall regional GDP growth of about 2% in 2021 on average (compared to ~4% in our baseline scenario).

Next to these key aspects, there are three further assumptions underlying our identification approach whose validity should be checked in complementary ex-post evaluations: First, in the specification of the regional growth model we assume a relatively fast input-output transmission channel. That is, BF funding is assumed to show up in regional GDP figures in the following year. Although we find empirical support for this lag structure in our sample, there are also good reasons to believe that it takes further time until R&I activities fully translate into economic output. In this sense, our predictions may actually underestimate the mid- to long-run impact of BF funding on the Finnish economy.

Second, we assume that national economic development exogenously affects regional GDP growth trends. While this is needed in our short-run forecasting system to capture the COVID-19 shock, alternative approaches using spatial rather than national-regional dependences should be additionally explored in the future (see Baltagi et al., 2014).

Third, there is the issue of causality. As we have already pointed out in the paper, we identify effects under the assumption of weak exogeneity of regressors (including BF funding). That is, that they are predetermined with respect to current regional GDP growth rates. While this may be sufficient to estimate robust conditional correlations at the regional level as a means to produce reliable short-run GDP growth forecasts, we cannot make strong statements about the causality between BF funding and economic growth. Future studies on the effectiveness of public R&I funding under COVID-19 should thus consider alternative ways to identify causal relations, preferably on the basis of micro data for firms and institutions receiving public R&I funding. Ultimately, there are many unknowns and uncertainty that make forecasting such a complex crisis as the COVID-19 pandemic has caused extremely challenging (Luo, 2021). We, though, hope that our ex-ante assessment may nonetheless provide valuable priors for this endeavor and support policy makers with highly needed information to assess and finetune their policies during the ongoing COVID-19 crisis and for preparing for future crises.

**Funding**

Business Finland [grant number: 34030/31/2020].

**Conflict of Interest**

While the research was funded by Business Finland, the funder did not influence the results of the research. The work was conducted independently without any pressure from the funder, or the steering group appointed by the funder

Figure A1: Map of Finnish regions (regional capitals in brackets)

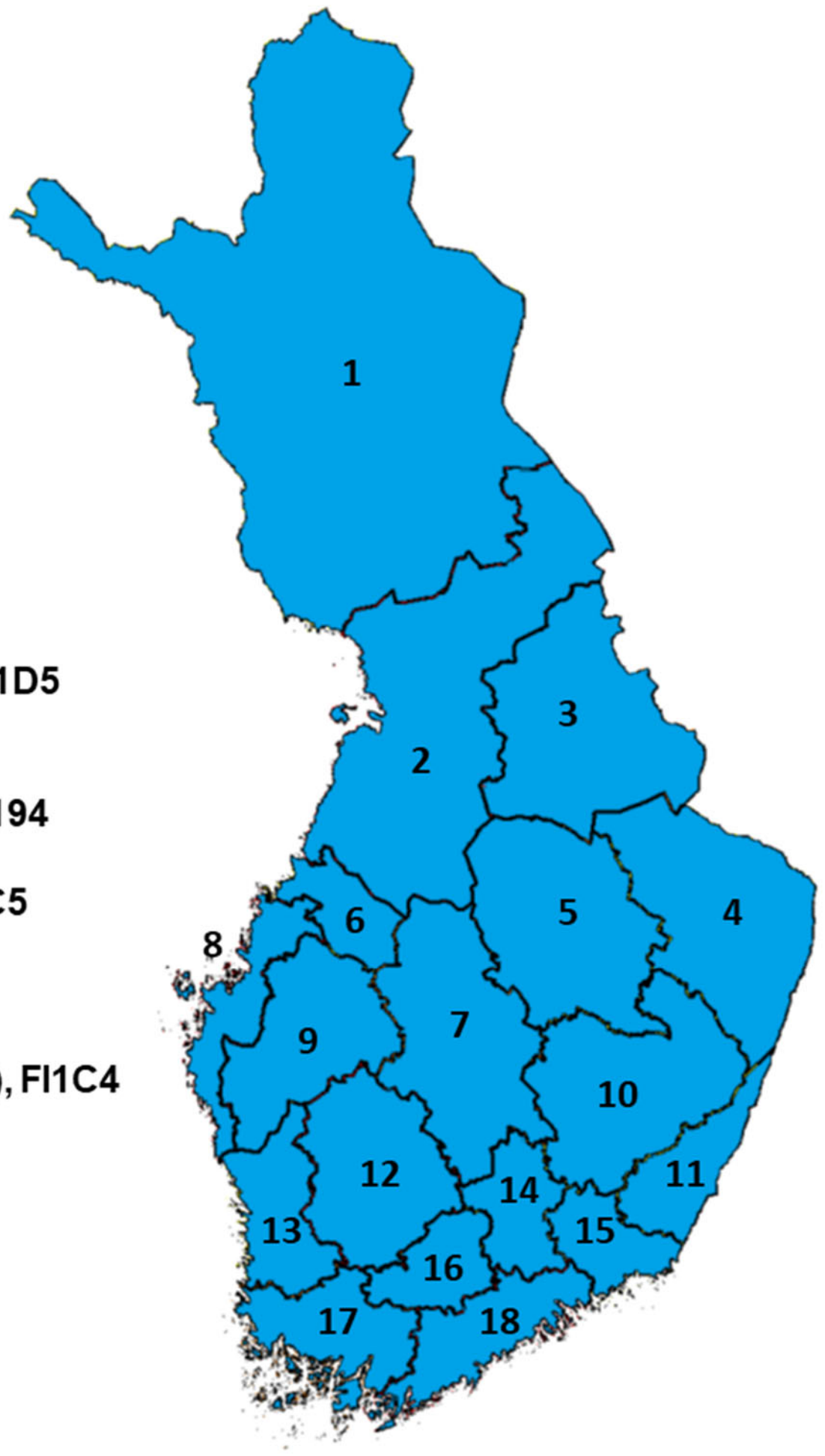

1. Lapland (Rovaniemi), FI1D7
2. North Ostrobothnia (Oulu), FI1D9
3. Kainuu (Kajaani), FI1D8
4. North Karelia (Joensuu), FI1D3
5. North Savo (Kuopio), FI1D2
6. Central Ostrobothnia (Kokkola), FI1D5
7. Central Finland (Jyväskylä), FI193
8. Ostrobothnia (Vaasa), FI195
9. South Ostrobothnia (Seinäjoki), FI194
10. South Savo (Mikkeli), FI1D1
11. South Karelia (Lappeenranta), FI1C5
12. Pirkanmaa (Tampere), FI197
13. Satakunta (Pori), FI196
14. Päijät-Häme (Lahti), FI1C3
15. Kymenlaakso (Kotka and Kouvola), FI1C4
16. Kanta-Häme (Hämeenlinna), FI1C2
17. Southwest Finland (Turku), FI1C1
18. Uusimaa (Helsinki), FI1B1

Figure A2: In- and out-of-sample forecasts for the individual NUTS-3 regions in Finland

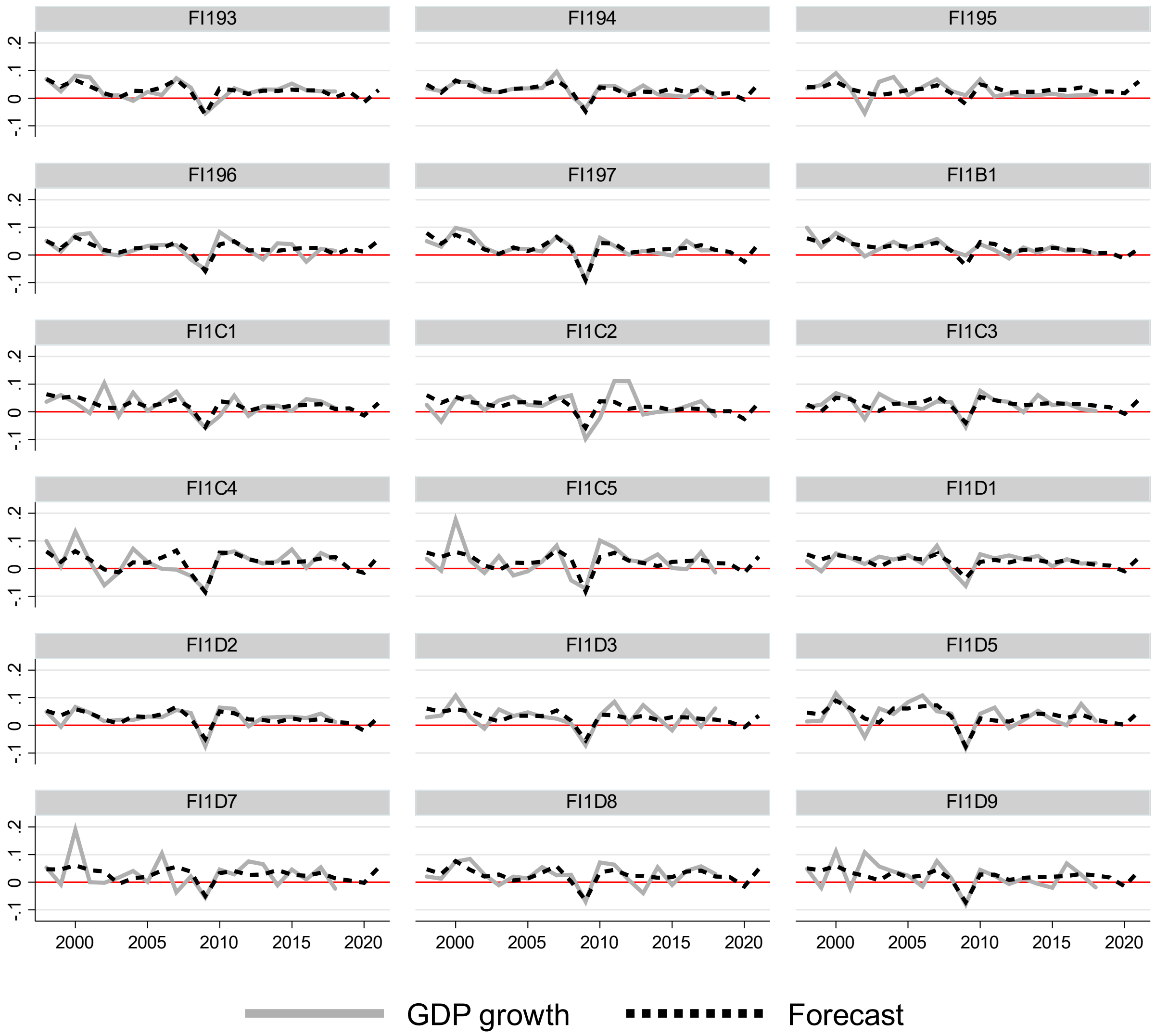

*Notes:* Own estimates based on parameters reported in Column (5) of Table 3 and forecast setup described in Table 2. National GDP growth in Finland in 2021 is assumed to be 1.8%. Region names associated with Finnish NUTS-3 codes are provided in Figure A1.